\documentclass[11pt]{article}
\usepackage{graphicx,url,epstopdf}

\begin{document}
\title{Transparent Covid-19 prediction} 
\author{Christoph Bandt\thanks{Institute of Mathematics, University of Greifswald, 17487 Greifswald, Germany, bandt@uni-greifswald.de} } 
\maketitle

\begin{abstract}
We present a very simple and transparent method to interpret time series of confirmed Covid-19 cases.  Roughly speaking, the analysis of weekly new infections for each day is a tool for the definition and early detection of the turning point of the epidemic. In Italy, the growth of Covid-19 activity was overcome one week ago, Austria and Switzerland followed suit. This note emphasizes the crucial delay of two weeks between infection and listing of infection in a central database. Our  estimates of the time course of Covid-19 activity and reproduction rates can reduce the information gap by several days.  We show that Spain and Germany are already beyond the turning point. In general the estimated reproduction rates become undercritical a short time after the main lockdown measures, and the Covid-19 activity starts to decrease afterwards. This note gives a very optimistic outlook for all regions which have taken strict containment measures.
\end{abstract}
 
\begin{figure}[h!t] 
\begin{center}
\includegraphics[width=0.99\textwidth]{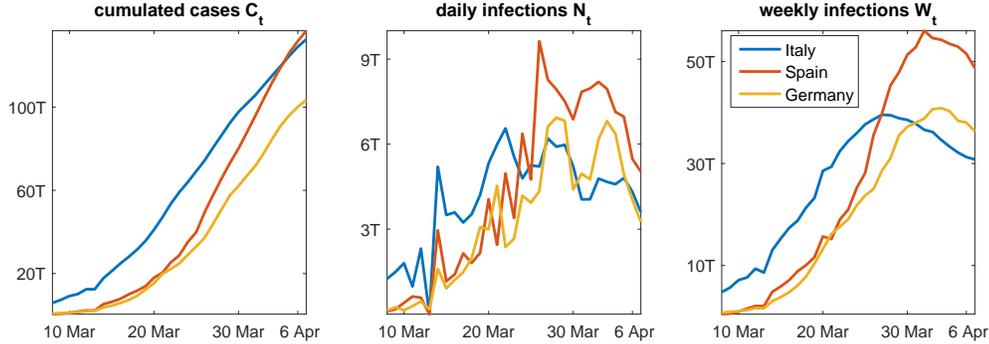} 
\end{center}
\caption{Overall number $C_t$ of confirmed cases, number $N_t$ of daily infections, and number $W_t$ of weekly infections for Italy, Spain and Germany, March 8 to April 7, 2020.}\label{fi1}
\end{figure}

\subparagraph{The data.}  To describe the global Covid-19 infections, Johns Hopkins University established the now famous database of cumulated numbers of confirmed cases, deaths and recoveries, for each day $t$ and each country or region \cite{JHU}. Here we explain how the series $C_t$ of confirmed infections for a certain region can be interpreted and used for prediction. The letter $t$ is commonly used  to denote either a date, or a number, like `day 10 from the beginning of observations'. If $t$ is today, then $t-1$ is yesterday, and $t-7$ is one week ago.
The sequence $C_1,C_2,...$ is usually drawn as a function of $t,$ as shown on the left of Figure \ref{fi1}. It is rather smooth and ever-increasing.  It is compared with exponential growth functions - a too pessimistic viewpoint since containment measures gradually diminish the growth. Another data series is given by the number of new infections on day $t,$  
\[ N_t= C_t-C_{t-1} \ .\]
This function illustrates the time course of infections much better. It is irregular since the data were collected under pressure. For instance, in the middle of Figure \ref{fi1} we see the value 4749 for Spain on 25 March followed by 9630 the day after. In $C_t$ this deviation is hardly noted.  France was not included in the picture because a still bigger outlier, 25615 on April 5 and 2886 on April 6, at present makes our analysis impossible.

\subparagraph{Basic idea.} Here we consider the numbers of weekly new infections
\[ W_t= C_t-C_{t-7}=N_t+N_{t-1}+...+N_{t-6}  \]
for each day $t.$ Although the $W_t$ are composed of past values, they have a greater predictive value than daily infections. The right of Figure \ref{fi1} clearly shows the course and the apex of the epidemic. Italy now is almost two weeks beyond the apex, Spain has followed, with a strong decrease of new infections, and Germany is also on the decreasing side. Below we prove that after lockdown conditions there must be a steady improvement during the first two weeks after the turning point - no matter whether lockdown is lifted or not. In this time the danger of infection is extremely low. And if an infection should happen, it will be included in $C_t$ only two weeks later. This is a very optimistic outlook, which with a time shift applies to all regions which have taken strict containment measures.

Note that the improvement is not visible in $C_t.$ Germany has a much smaller total number than Italy, and the same slope, but it reached the turning point several days later.  A conventional analysis would emphasize that Spain has largest slope and has just overtaken Italy in the total number of cases, so it should be in bad shape. However, weekly numbers show that Spain obviously is improving fast.

Weekly numbers are appropriate for Covid-19 analysis 
for several reasons. A moving sum is widely used as a smoothing filter of data \cite{Shum} so $W_t$ is much more regular than $N_t.$ 
Seen as a difference filter, the $W_t$ will also damp weekly periods in the data which often reflect the mechanism of health administration. 
The most important and specific argument is that $W_t$ can be considered as a proxy of the \emph{infection potential} or \emph{Covid-19 activity} $I.$ We are interested in the time course of the activity, not in its actual size. We neglect the uncovered asymptomatic individuals by assuming that the infections caused by them are proportional to the infections caused by positively tested persons.  

Many experts accept the rule that positively tested patients remain contagious for 7 days. See the Covid-19 activity maps in \cite{RKI}, for instance.  
So we just subtract from the overall number $C_t$ those cases which have become unable to spread the infection. Thus $W_t$ models the danger for the public. We want to know how this quantity changes in time. However, the most actual numbers $W_t$ give information only on the danger two weeks ago. Formally, $W_t$ is an estimate of $I_{t-14},$ not of $I_t.$ It is crucial to understand completely this information gap of Covid-19. A detailed discussion is given below.   

\begin{figure}[h!] 
\begin{center}
\includegraphics[width=0.8\textwidth]{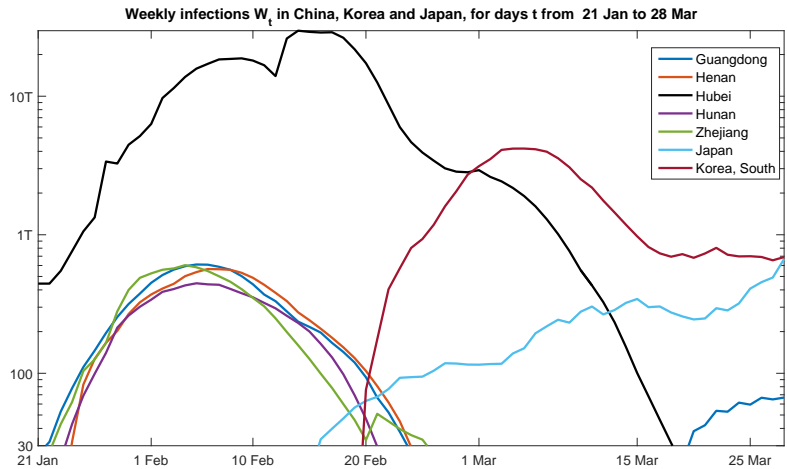} 
\end{center}
\caption{Development of weekly infection numbers $W_t$ for each day between 21 January and 28 March 2020 in China, South Korea, and Japan. The logarithmic scale makes it possible to show the Wuhan province and Chinese provinces with small number of infections in one figure. All regions except Japan go through a parabolic turning point. At the end of March, new infections in Guangdong have risen again above the minimum of 30.}\label{fi2}
\end{figure}

The time for which a patient can infect others varies between 7 and 14 days for moderate cases according to \cite{ecdc}, and this is why quarantine should last at least two weeks. The light cases which are not tested also cause infecitons. For the statistics, we take one week as a kind of mean value. We may also pretend that the time varies between 5 and 9 days, and replace $C_{t-7}$ in the definition of $W_t$ by the average $\frac{1}{5}(C_{t-5}+C_{t-6}+...+C_{t-9}).$ The resulting smoothed weekly infections were used only for Figure \ref{fi2} to obtain smooth curves.  Other coefficients would be possible, but we want to keep things transparent. 

In contrast to simulation studies based on models, like \cite{rep9,rep13}, our approach is strictly data-driven. We have only 20 up to 50 hastily collected daily numbers for the evaluation of a very complex process in a whole country. We take these data as they are, and adapt our methods to make the best of them. To fit  models to these data, many additional assumptions would be necessary. 
Our calculations can be checked and modified by everyone who has a computer and a software for calculation like Excel or R, or Matlab, which was used here. 

\subparagraph{The Chinese example.} 
China was the first country to go through the Covid-19 epidemic. Figure \ref{fi2} shows the development of weekly infection numbers $W_t$ in Hubei with capital Wuhan,  four other Chinese provinces, South Korea, and Japan. This picture conveys an optimistic message. We see that on the right side of the maximum the function  decreases exponentially fast, only slightly flatter than on the increasing side. The time in Figure \ref{fi2} starts just before 23 January when strict lockdown measures in Wuhan were implemented by the Chinese government. Note that the scale of $W_t$ is logarithmic between 30 and 40000. The statistics in Wuhan could be the best with so many cases but on Feb 12 the criteria for confirmed cases were changed and 15000 cases added, after 0 cases the day before. This explains why the curves for the other provinces look more regular, like parabolas. On linear scale, they would be bell-shaped. 

In the provinces with relatively few cases, as well as in Korea, it took a bit more than two weeks to reach the apex of Covid-19 activity. In Wuhan, with tens of thousands of active infections, four weeks were necessary, starting from lockdown. European countries are going through a similar scenario, and Italy was first to  overcome the apex in Figure \ref{fi1}. 

Figure \ref{fi2} shows that Covid-19 activity hardly disappears completely. Even with very strong restrictions in Wuhan, it stayed on an intermediate level of 3000 for several days, and then went down to almost zero. In Guangdong the infection potential has again risen above 30. In South Korea, where no emergency measures were implemented, the infection potential decreased up to 700 and then stayed constant. It is not easy to manage Covid-19 activity of such size, but it seems possible for a large and technologically developed country. Japan has so far succeeded in keeping the infection potential low, though it crossed the Korean level at the end of March.   

\subparagraph{Delays.} We must understand that the infections counted today did actually happen two up to four weeks ago, as indicated in the following scheme. 
\vspace{2ex}

$\bullet \rule[1mm]{20mm}{1mm}\bullet \rule[1mm]{20mm}{1mm}
\bullet \rule[1mm]{20mm}{1mm} \bullet \rule[1mm]{20mm}{1mm}
\bullet \rule[1mm]{38mm}{1mm} \bullet$ \\
infection\hspace{5mm}symptoms\hspace{5mm}test of patient\hspace{5mm}
lab result\hspace{5mm}local report\hspace{3mm}report to region/government
\vspace{1ex}

The delay between infection and appearance of symptoms, known as incubation time, is beyond human control. It is said to be between 2 and 12 days, with 5-6 days in average \cite{WHO,ecdc}. More details and references can be found at \cite{worldo}.
Depending on the strength of symptoms and the time, for example Saturday night, the patient will decide to undertake a test. To this end, a doctor has to be consulted who performs the test or certifies the need of testing at another place. The test material is transported to a lab and evaluated. The lab reports to the doctor and/or a local authority. A file is created and a report sent to regional and federal authorities. 

It seems obvious that the total time between an infection and its reporting to the government is rarely below 10 days.  In Europe a realistic estimate of the average is between two and three weeks. There are several moments of hesitation. A patient hesitates to visit the doctor. A local officer hesitates to report a bad case because the press may meddle into his actions. Moreover, a health administration trimmed for cost-efficiency has no reserves and slows down under pressure. Finally, officials may underestimate the importance of fast information.

All delays are of course random variables, and beside mean values, variances are important. Careful analysis of delays is crucial since they dramatically influence political decisions. All over the world the danger of an epidemic was initially underestimated because of late information. This is a main reason for the current crisis. A simulation study by Ferguson et al. \cite{rep9} came to the conclusion that 20 million people will die if governments do not act firmly and rapidly enough.

\subparagraph{The dilemma of politics.} 
When the danger was realized, harsh measures were taken: borders and schools closed, mobility restricted, emergency laws activated. However, even the most drastic measures cannot avoid an infection which already took place. We have to wait two weeks to see the very first effects of containment measures.
This is an unprecedented and extremely difficult situation for politicians, for the people, and for the media who live from news. Nobody is used to wait two weeks when the situation gets worse. 

Pressed by the media, politicians are tempted to order more and more restrictions, no matter how expensive or inefficient. When the situation improves, they can claim success. Otherwise, the measures were necessary anyway.

Imagine a hunter fighting against a raging tiger, while he always sees only the tiger's position 14 seconds ago.  He will take a very heavy machine gun and shoot all his ammunition, destroying the forest and killing friendly animals, since he is never sure that the tiger was hit. 

The disadvantage of delays in the period of infection growth turns into an advantage when the apex is overcome. For two weeks the situation will constantly improve, no matter what we do. Even if we become careless, we shall notice this only two weeks later. Nevertheless, at the start of improvement we can immediately lift all restrictions of the previous two weeks, because they did not contribute to the change. They simply had no chance to show an effect.

\subparagraph{Five European countries.} While the epidemic in the USA and in the United Kingdom is still in a growing stage, some European nations have been in the public focus for almost four weeks now. Since in Wuhan the time from lockdown to the apex was about four weeks, did they also arrive at the turning point?  

We now look on Figure \ref{fi1} on logarithmic scale for the last two weeks.  
On the one hand, logarithms allow to show small and big numbers in one picture, as in Figure \ref{fi2}. So we can include Switzerland and Austria which both reached the apex one day after Italy. On the other hand, logarithms provide the appropriate scale for multiplicative processes where day-to-day differences are terribly growing but day-to-day quotients are slowly varying. 

\begin{figure}[h!t] 
\begin{center}
\includegraphics[width=0.99\textwidth]{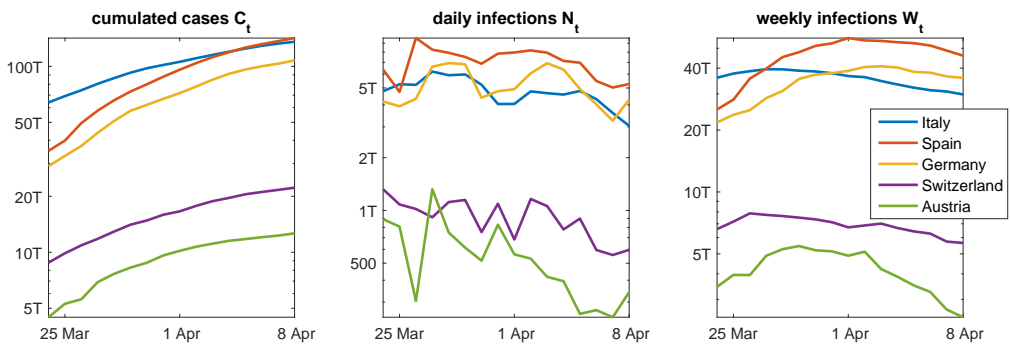} 
\end{center}
\caption{Confirmed cases $C_t,$ daily infections $N_t,$ and weekly infections $W_t$ for Italy, Spain, Germany, Switzerland, and Austria, March 18 to April 5, 2020.}\label{fi3}
\end{figure}  

The function $C_t$ becomes concave on logarithmic scale. That is, the function is still increasing, but its slope decreases with time. Note that the graph of an exponential function on logarithmic scale is a line. The decrease of slope in the graphs of $C_t$ vaguely illustrates that containment measures had an effect. 

Mathematicians like to find change points in time series, and certainly politicians would like to see a sudden change which can be identified as effect of a certain lockdown measure. Unfortunately, due to the big variance of delays we can only see gradual decrease of the slope. More detailed data are necessary to specify the effect of measures.

Both $C_t$ and $N_t$ give little information on differences between countries. Only  $W_t$ shows such differences. While Italy, Switzerland and Austria reached the apex almost together, Spain and Germany came later. The weekly infection numbers of Italy have steadily decreased for 13 consecutive days now. In Spain and Germany these numbers decreased for 6 consecutive days. For Austria and Switzerland the decrease of $W_t$ was interrupted. The reason is that $N_t$ for small countries is subject to greater statistical variation. (For a Poisson random variable, the 95\%  confidence radius is $2\sqrt{N}.$ Since Covid-19 cases come in clusters, we should take twice that value, so that $N=500$ has a statistical error $\pm 100$ or 20\%   while $N=5000$ has an error $\pm 300$ or  8\% .)

The decrease of $W_t$ in Austria is fast, as for Chinese provinces with small infection numbers in Figure \ref{fi2}. The decrease of $W_t$ for Italy, Spain and Germany is much slower, like for Wuhan in Figure \ref{fi2}. We can suspect that today, two weeks later, the values are actually on the level of Wuhan at 1 March. 

On the whole, the decrease of $W_t$ on logarithmic scale does not look so convincing. Is it possible that the values will rise again?  We have to study the infection process from another point of view.

\subparagraph{The reproduction rate.} 
A basic parameter of epidemiology is the reproduction rate $R$ of a disease which says how many other persons will be infected by a sick individual. If it is larger 1, the epidemics will spread exponentially. If $R<1$ the epidemic will end soon. For Covid-19, estimates of $R$ vary between 1.4 and 6.5, see \cite{repro}.
The scenarios of \cite{rep9, rep13} were determined from mathematical models which are complex improvements of so-called SIR models, They have  convinced politicians that strong containment measures are unavoidable and save lifes. 

Now we are in the phase of lockdown which is difficult to describe by such models. The spatial aspect has to be taken into account. A first spatial infection model was the contact process \cite{Lig}, today one would use random geometric graphs  \cite{Pen}. Much more and better data are needed to fit parameters of such models to Covid-19 reality.

So we shall try to estimate reproduction for each country from the small and rough data series $C_t.$ This works surprisingly well. We postulate that the new infections $N_t$ come from the yesterday's infection potential $W_{t-1}.$  That is quite realistic: on the one hand the infection took place several days ago, but on the other hand the person who caused today's infection would himself remain infectious for a few more days. Thus a \emph{daily} reproduction rate can be estimated as the quotient $q_t= N_t/W_{t-1}.$
 
Unfortunately, calculation of quotients will not at all work with irregular data. We must care for smoothing in both numerator and denominator.  Instead of $N_t$ we take the average $(N_t+N_{t-1})/2 =(C_t-C_{t-2})/2.$ The denominator now must also be stretched over two time points. We replace $W_{t-1}=C_{t-1}-C_{t-8}$ by $(C_{t-1}+2C_{t-2})/3 -C_{t-11}.$ We neglected smoothing at $t-11$ since we worked on the increasing branch of $W_t$ and wanted predict the turning point. 
The resulting quotients were still not smooth enough, and were smoothed by a simple binomial filter. This gives the formula
\[  q_t=\frac{p_{t-1}+2p_t+p_{t+1}}{4}\quad\mbox{ with } \quad
p_t=\frac12\frac{C_t-C_{t-2}}{(C_{t-1}+2C_{t-2})/3 -C_{t-11}}.\]

\begin{figure}[h!t] 
\begin{center}
\includegraphics[width=0.9\textwidth]{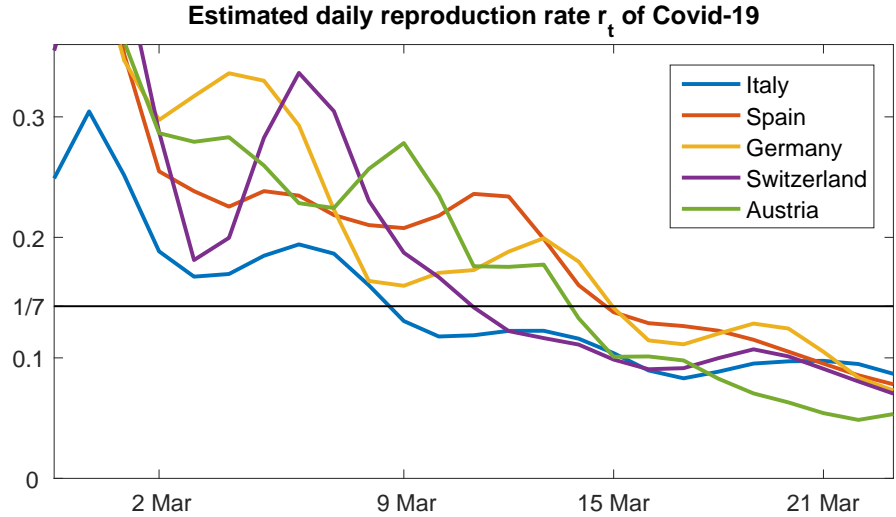} 
\end{center}
\caption{The estimated reproduction rate $r_t.$ }\label{fi4}
\end{figure}  

Note that the smoothing results in one and a half day loss of actuality. So because of the delay of 14 days, the number $q_t$ is an estimate for the daily reproduction rate of the disease at time $t-15.5.$ These daily reproduction rates $r_t$ are shown
in Figure \ref{fi4}. The time is now the time of real infection, not the time of observation.

When a positive case remains infectious for seven days in the average, then its total reproduction rate is seven times larger than the daily one. So the critical value of the daily reproduction rate is $1/7$ instead of one. In Figure \ref{fi4}, the critical value is indicated by a black line.  It turns out that the curves of all five countries hit this critical line, in almost the same order as the apex of estimated Covid-19 activity was reached. At the end of the time period, on March 23, Austria had the smallest reproduction rate, less than half of the critical value. The other four countries have almost the same reproduction rates, all far lower than the critical value. From this point of view the infection process will die out, according to the paradigm of epidemiology. This holds under constant lockdown conditions which have been in force for two more weeks now.  Thus also from this point of view, the actual danger of infection today, at April 8, is very low.  

\subparagraph{Conclusion.} 
Finally, let us compare the following dates in Table \ref{ta}. \\
$\bullet\ $ The day of maximal Covid-19 activity, which is two weeks before our observed maximum point of $W_t$ in Figure \ref{fi3}.\\
$\bullet\ $ The day in Figure \ref{fi4} where the estimated daily reproduction rate crosses the critical line.\\
$\bullet\ $ The day of stay-at-home order in the respective country. 

The lockdown measures in different countries differ a lot and cannot be compared so easily. However, in each country there was a definite speech of the head of government with a strict stay-at-home order. There were less restrictive containment measures before, and there were further restrictions after that date.
We refer to \cite{rep13} for a comprehensive comparison of dates.
 
In Italy first the northern regions with many infections were locked. On March 5 events were forbidden and schools closed, followed by complete lockdown on March 11. In Germany some federal states took their specific measures earlier, schools were closed from March 15, events completely forbidden on 20, and the lockdown came on March 22. \cite{RKI}. Everywhere some measures were announced in advance and took place one or more days later.

\begin{table}[h]
\centering
\begin{tabular}{|c||c|c|c|c|c|} \hline
country&\ Italy\ &\ Spain\ &Germany&Switzerland&Austria\\ \hline 
estimated apex&13&18&20&12?&15\\ \hline 
hitting critical line&9&15&15&14&11\\ \hline
stay-at-home order&11&14&22&16&20\\ \hline
\end{tabular}
\caption{Day of March for turn of Covid-19 activity, compared with day where  reproduction becomes undercritical, and day of the strong lockdown}\label{ta}
\end{table}

The essential observation of our study is: the lockdown measures had an immediate impact. The reproduction rate crossed the critical line almost instantly, sometimes even before the main lockdown speech, and the apex of Covid-19 activity was reached very few days later. The tiger was hit with the first or second shot. Politicians could rely on the people's cooperation and willingness to fight the disease.  Further restrictions after the main speech could not show effects, however, since the virus is on very low level two weeks after their implementation. Due to the information gap, the danger was overestimated, and more collateral damage done than needed.

\subparagraph{The necessity to minimize the information gap.} 
Now restrictions have to be carefully lifted. The tiger is not dead, it will come back. Again we may underestimate the danger - not in the very moment, but a few weeks ahead.

At this point, it is absolutely crucial to minimize the information delay. New monitoring systems must be built up. They cannot be run by the local health administrations who do a great job, have enough work and an established workflow. There must be parallel lines of information.  One cheap improvement is the establishment of an immediate daily report on  numbers of positive and negative tests from every lab to the health ministry.  In Germany, this could probably gain three days for government decisions.  An even earlier although not definite information would be provided by a central 24/7 hotline where everybody can apply for a date in a test center, and where basic information on symptoms and contact to infected people can be gathered just after the incubation time. Different lines of data need to be aggregated, for which experts have to be hired. This is essential for getting through to the time when a vaccine will be available.

\bibliographystyle{plain} 
\bibliography{copre1}

\end{document}